\documentclass[12pt]{article}
\usepackage{amsfonts}
\usepackage{amssymb}
\usepackage{cite}

\newcommand{\no}{\nonumber}
\newcommand{\Eqn}[1]{&\hspace{-0.5em}#1\hspace{-0.5em}&}

\newcommand{\tfrac}[2]{{\textstyle\frac{#1}{#2}}}

\newcommand{\hK}{{\hat{K}}}

\newcommand{\hrho}{{\hat{\rho}}}

\newcommand{\alg}[1]{\mathfrak{#1}}

\def\pint#1 {- \!\!\!\!\!\!\!\! \,\int_{#1}}

\def\ni       {\noindent}

\def\semiket#1  { \, #1 \, \rangle \, }

\def\abs#1      {  \, \vert #1 \vert \,   }
\def\Im#1    { \, {\rm Im } \, #1  }
\def\Re#1    { \, {\rm Re}  \, #1  }

\def\binom#1#2 { \vecii{ {}_{#1} }{\raisebox{.5ex}{$ {}^{#2} $}} }
\def\sqbinom#1#2 { \Bigl(\begin{array}{c} {}_{#1}
                       \\ \raisebox{.5ex}{${}^{#2}$} \end{array}\Bigr)^2  }

\def\r12    {\frac{r_1}{r_2}}

\def\vecii#1#2      {  { #1 \choose #2 }  }
\def\veciii#1#2#3   {  \left(\begin{array}{c}#1\\#2\\#3\end{array}\right)  }
\def\matrixii#1#2#3#4            {  \Bigl( \begin{array}{cc}#1&#2\\#3&#4
                                       \end{array} \Bigr) }
\def\matrixiii#1#2#3#4#5#6#7#8#9 {  \left(\begin{array}{ccc}#1&#2&#3\\
                                     #4&#5&#6\\#7&#8&#9\end{array}\right)  }
\def\eqb         {  \begin{eqnarray}  }
\def\eqe           {  \end{eqnarray}  }

\def\csectionast#1    { \begin{center}
    {\large\bf #1  }   \end{center} \par \bigskip}
\def\titleandfile#1#2   {  \begin{center}{\large\bf #1}\end{center}
                            \par\begin{flushright} #2 \end{flushright} 
                            \par \begin{flushright} \today \end{flushright}}

\renewcommand{\thefootnote}{\fnsymbol{footnote}}

\begin{document}
%
\def\papertitlepage{\baselineskip 3.5ex \thispagestyle{empty}}
\def\preprinumber#1#2#3{\hfill \begin{minipage}{4.2cm}  #1
              \par\noindent #2
              \par\noindent #3
             \end{minipage}}
\renewcommand{\thefootnote}{\fnsymbol{footnote}}
%
%
\papertitlepage
\setcounter{page}{0}
\preprinumber{}{UTHEP-540}{hep-th/0703177}
\baselineskip 0.8cm
\vspace*{2.0cm}
\begin{center}
{\large\bf\mathversion{bold}
On the origin of the dressing phase 
in ${\cal N}\!=4$ Super Yang--Mills}
\end{center}
\vskip 4ex
\baselineskip 1.0cm
\begin{center}
        { Kazuhiro ~Sakai\footnote[2]{\tt sakai@phys-h.keio.ac.jp}, } \\
 \vskip -1ex
    {\it Department of Physics, Keio University}
 \vskip -2ex   
    {\it Hiyoshi, Yokohama 223-8521, Japan} \\
 
 \vskip 2ex
     { Yuji  ~Satoh\footnote[3]{\tt ysatoh@het.ph.tsukuba.ac.jp}}  \\
 \vskip -1ex
    {\it Institute of Physics, University of Tsukuba} \\
 \vskip -2ex
   {\it Tsukuba, Ibaraki 305-8571, Japan}

\end{center}
\vskip 13ex
%
\baselineskip=3.5ex
\begin{center} {\bf Abstract} \end{center}

We derive the phase factor proposed by Beisert, Eden and Staudacher 
for the S-matrix of planar ${\cal N}=4$ Super Yang--Mills,
from  the all-loop Bethe ansatz equations without the dressing factor.
We identify a configuration of the Bethe roots, 
from which the closed integral formula of 
the phase factor is reproduced in the thermodynamic limit.
This suggests that our configuration describes the ``physical vacuum'' 
in the sense that the dressing phase is nothing but  the effective phase 
for the scattering of fundamental excitations above this vacuum, 
providing an interesting clue to the physical origin of the dressing phase. 

\vskip 2ex
%
%
%
%
%
\vspace*{\fill}
\ni
March 2007
\newpage
\renewcommand{\thefootnote}{\arabic{footnote}}
\setcounter{footnote}{0}
\setcounter{section}{0}
\baselineskip = 3.3ex
\pagestyle{plain}

Integrability has become of increasing importance
in the study of ${\cal N}=4$ Super Yang--Mills (SYM)
and of the dual superstrings in $AdS_5\times S^5$.
The spectral problem of the dilatation operator at one loop
was identified with that of a conventional integrable
spin-chain \cite{Minahan:2002ve,Beisert:2003yb},
which can be systematically solved by using Bethe ansatz.
Integrability beyond one loop has also been extensively
studied and, in particular, 
the all-loop Bethe ansatz equations were postulated
\cite{Beisert:2005fw}.
Note, however, that the spin-chain picture does not fully apply
at higher loops due to several new features
yet unknown in the field of integrable models,
such as length fluctuation.
Nevertheless, conventional integrability revives
by converting the picture into a particle model,
at least in the limit of infinite length of operators,  
or the large-spin limit \cite{Berenstein:2002jq,Gubser:2002tv,Frolov:2003qc}.
Asymptotic particle states were realized
in terms of SYM operators \cite{Staudacher:2004tk}.
It is expected that
they exhibit the factorized scattering property and thus
all the multi-particle scattering processes are governed
by the elementary two-particle S-matrix.
This S-matrix was determined up to an overall scalar factor
by purely algebraic consideration of the centrally extended
$\alg{su}(2|2)$ symmetry \cite{Beisert:2005tm}
and further algebraic aspects have been investigated
\cite{Beisert:2006qh,Arutyunov:2006yd,Martins:2007hb}.

As is expected from the AdS/CFT correspondence,
this S-matrix with a pair of the $\alg{su}(2|2)$
symmetries also emerges
on the string theory side 
\cite{Arutyunov:2006ak,Klose:2006zd,Arutyunov:2006yd,Klose:2007wq}.
The choice of the gauge breaks the conformal invariance in
two dimensions and one obtains a massive worldsheet theory,
where S-matrix is naturally defined as the scattering of
elementary excitations.
As the symmetry completely constrains the form
of the matrix, what is left to be determined
is again the overall scalar factor.

The determination of the scalar factor,
as a function of two momenta and the coupling,
is important in two aspects: Firstly, it is the last missing
element for the systematic construction of the 
spectrum of the scaling dimension/energy
on the Yang--Mills/string side.
Secondly, identification of the scalar factors
on both sides serves as a strong quantitative check
of the AdS/CFT correspondence.

The form of the scalar factor was first studied
on the string side, based on the data of
classical string spectrum \cite{Arutyunov:2004vx}.
Succeedingly $1/\sqrt{\lambda}$ corrections were analyzed
\cite{Hernandez:2006tk,Arutyunov:2006iu,Freyhult:2006vr}
and an all-order form was postulated \cite{Beisert:2006ib}.
This form was shown \cite{Arutyunov:2006iu,Beisert:2006ib}
to be consistent with
the constraint from the crossing symmetry \cite{Janik:2006dc}.
On the other hand, the form of scalar factor
was rather obscure on the Yang--Mills side,
since it stays trivial up to three loops.
However, it turned out to deviate from the unity
at four loops \cite{Bern:2006ew,Cachazo:2006az}.
Meanwhile,  Beisert, Eden and Staudacher managed to construct
a closed integral formula \cite{Beisert:2006ez}
consistent with the above four-loop result
as well as a sort of analytic continuation of
the proposal on the string side \cite{Beisert:2006ib}.
The integral formula is highly intricate,
but does not seem totally hopeless to handle analytically.
We refer to a reference \cite{Dorey:2007xn} for recent investigations.

In this short article, we present the derivation
of the integral formula purely within the framework
of quantum integrable models.
Our result is of conceptual importance,
since it would indicate that even the scalar factor
has no room to consult model-specific degrees of freedom.
The integrable structure together with the $\alg{su}(2|2)$ 
symmetry would completely determine the S-matrix
without knowing which side of the AdS/CFT correspondence 
we are looking at.

Before getting into our computation,
we would like to remind the reader of the
derivation of the Zamolodchikovs' S-matrix \cite{Zamolodchikov:1978xm}.
This S-matrix describes the scattering of
elementary particles of the principal chiral field model.
It was originally determined by imposing three conditions:
unitarity, associativity (Yang--Baxter equation),
and crossing symmetry.
The first two determine the form of R-matrix,
while the last constrains the overall scalar factor
up to the CDD ambiguity.
The S-matrix can also be derived by direct computations
\cite{Korepin:1979qq,Faddeev:1981ft,Andrei:1983cb}.
In this case, the starting point is bare Bethe ansatz equations
derived from the R-matrix.
The physical S-matrix is realized as
the scattering matrix of excitations
over the non-trivial physical vacuum state, which 
is built on the bare vacuum state by acting with bare Bethe roots
filling up the Dirac sea.
Scattering of fundamental excitations above the Dirac sea
acquires a phase shift due to the interaction with
those background Bethe roots.
The phase shift then turns into the
scalar factor in front of the bare R-matrix,
giving the Zamolodchikovs' S-matrix.

In what follows we consider an analog of this derivation.
Along this line, possibilities of deriving the scalar
(dressing) factor for planar ${\cal N}=4$ Super Yang--Mills
have been discussed in a recent work \cite{Rej:2007vm}.
We refer to a work \cite{Gromov:2006cq} for
a somewhat similar approach.

\par\bigskip
Our starting point is the all-loop Bethe ansatz
equations \cite{Beisert:2005fw} without the dressing factor.
Most generally the Bethe ansatz equations consist of seven sets of
equations. For our purpose,
we set the number of Bethe roots as\footnote{
This set of occupation numbers is not allowed at one loop,
but in the present case
it is consistent with the bound from
the consistency of
the nested Bethe ansatz
$K_2 \le K_1+K_3 \le K_4 \ge K_5+K_7 \ge K_6$,
as long as $K_4\ge 2M$ is satisfied \cite{Martins:2007hb}.
(We would like to thank A.~Rej, M.~Staudacher and S.~Zieme
for discussions making us clarify this point.)
The numbers of Bethe roots are determined so that the corresponding
state is neutral under the pair of $su(2|2)$ symmetries. See for details \cite{Sakai:2007ie}.}
\begin{equation}
(K_1,\ldots,K_7)=(2M,M,0,K_4,0,M,2M).
\end{equation}
Bethe roots $u_{3,k}\,u_{5,k}$ as well as equations for them
are absent in this case.
Throughout this article we consider configurations of Bethe roots
symmetric with respect to the interchange of the two $\alg{su}(2|2)$
sectors: distribution of roots $u_{1,k},u_{2,k}$ is just the same as
that of $u_{7,k},u_{6,k}$, respectively. We thus omit the
equations for $u_{1,k},u_{2,k}$ below.
After all, we are left with the following reduced sets of
equations
\begin{eqnarray}
\label{BAE4}
\left(\frac{x^+_{4,k}}{x^-_{4,k}}\right)^L
\Eqn{=}\prod_{j\ne k}^{K_4}\frac{u_{4,k}-u_{4,j}+i}{u_{4,k}-u_{4,j}-i}
\prod_{j=1}^{2M}\frac{1-g^2/x^-_{4,k}\,x_{1,j}}{1-g^2/x^+_{4,k}\,x_{1,j}}
\prod_{j=1}^{2M}\frac{1-g^2/x^-_{4,k}\,x_{7,j}}{1-g^2/x^+_{4,k}\,x_{7,j}}\,,\\
1\Eqn{=}\label{BAE6}
\prod_{j\ne 1}^{M}\frac{u_{6,k}-u_{6,j}-i}{u_{6,k}-u_{6,j}+i}
\prod_{j=1}^{2M}\frac{u_{6,k}-u_{7,j}+i/2}{u_{6,k}-u_{7,j}-i/2}\,,\\
1\Eqn{=}\label{BAE7}
\prod_{j=1}^{M}\frac{u_{7,k}-u_{6,j}+i/2}{u_{7,k}-u_{6,j}-i/2}
\prod_{j=1}^{K_4}\frac{1-g^2/x_{7,k}\,x^+_{4,j}}{1-g^2/x_{7,k}\,x^-_{4,j}}\,.
\end{eqnarray}
It turns out that these equations describe, among others, a generalization
of the anti-ferromagnetic state of the $\alg{su}(2)$ Heisenberg spin-chain.
We consider the case where both $M$ and $K_4$ are of order $L$,
which will be sent to infinity in the thermodynamic limit.
We follow the standard parametrization that
rapidity variables $x$, $u$ are related by
\begin{equation}
x^\pm(u)=x(u\pm\tfrac{i}{2}),\quad
x(u)=\frac{u}{2}\left(1+\sqrt{1-4g^2/u^2}\right),
\end{equation}
and
\begin{equation}
g=\frac{\sqrt{\lambda}}{4\pi}
\end{equation}
is the normalized coupling constant.

We first recall that neighboring roots 
$u_{6,k}$ and $u_{7,j}$ attract each other
and may form bound states called stacks \cite{Beisert:2005di}.
Here we consider a particular type of stacks
studied in \cite{Rej:2007vm} that
every bosonic root $u_{6,k}$ is combined with
a 2-string of fermionic roots $u_{7,k}$.
The center of the 2-string coincides with the bosonic root
up to ${\cal O}(\frac{1}{L})$ correction.
With appropriate ordering of Bethe roots, one can express
the present formation of stacks as
\begin{equation}\label{stackrel}
u_{7,2k-1}\approx u_{6,k}+\tfrac{i}{2},\quad
u_{7,2k}\approx u_{6,k}-\tfrac{i}{2},
\quad \mbox{for}\quad {k=1,\ldots,M},
\end{equation}
where we let $\approx$ denote equality
up to ${\cal O}(\frac{1}{L})$ correction.
After substituting (\ref{stackrel}),
(\ref{BAE4}) read
\begin{eqnarray}\label{BAE4bis}
\left(\frac{x^+_{4,k}}{x^-_{4,k}}\right)^L
\Eqn{\approx}\prod_{j\ne k}^{K_4}\frac{u_{4,k}-u_{4,j}+i}{u_{4,k}-u_{4,j}-i}
\prod_{j=1}^{M}\frac{1-g^2/x^-_{4,k}\,x^+_{2,j}}{1-g^2/x^+_{4,k}\,x^+_{2,j}}
\prod_{j=1}^{M}\frac{1-g^2/x^-_{4,k}\,x^-_{2,j}}{1-g^2/x^+_{4,k}\,x^-_{2,j}}
\no\\
&&\hspace{6.79em}\times
\prod_{j=1}^{M}\frac{1-g^2/x^-_{4,k}\,x^+_{6,j}}{1-g^2/x^+_{4,k}\,x^+_{6,j}}
\prod_{j=1}^{M}\frac{1-g^2/x^-_{4,k}\,x^-_{6,j}}{1-g^2/x^+_{4,k}\,x^-_{6,j}}
\,,
\end{eqnarray}
while (\ref{BAE6}) are identically satisfied\footnote{ 
There appear seemingly indeterminate factors $0/0$
at the leading order of the large $L$ approximation.
This indeterminateness is resolved
if one takes account of the $1/L$ correction.
The requirement for the correction is that
$u_{7,2k-1}-u_{6,k} = i/2 + \epsilon_k$ and
$u_{7,2k}-u_{6,k} =  -i/2 - \epsilon_k$,
with $\epsilon_k={\cal O}(1/L)$.
Such adjustments are possible stack by stack.}.
Equations (\ref{BAE7}) split up into two kinds, namely, for
$u_{7,2k-1}$ and for $u_{7,2k}$.
Multiplying the former by the latter for the same $k$,
we obtain a set of center equations
\begin{equation}\label{BAE7center}
1\approx
\prod_{j\ne k}^{M}\frac{u_{6,k}-u_{6,j}+i}{u_{6,k}-u_{6,j}-i}
\prod_{j=1}^{K_4}\frac{1-g^2/x^+_{6,k}\,x^+_{4,j}}{1-g^2/x^+_{6,k}\,x^-_{4,j}}
\,\frac{1-g^2/x^-_{6,k}\,x^+_{4,j}}{1-g^2/x^-_{6,k}\,x^-_{4,j}}\,.
\end{equation}

Let us next consider the thermodynamic limit
$L\to\infty$ of these effective equations.
We are looking for a solution analogous to the anti-ferromagnetic state
of spin-chains, for which all the Bethe roots sit along the real axis
with consecutive mode numbers.
For such a solution we may well assume
the distribution of roots to be symmetric under $u\mapsto -u$.
For the sake of simplicity we set $K_4=L/2$,
which may be the possible maximal
number for the real roots $u_{4,k}$.
The number of stacks $M$ should also be fixed\footnote{
$M$ may be fixed so that the vacuum is
maximally neutral with respect to the global symmetry.
For $K_4 = 2M = L/2$, global charges listed in
\cite{Beisert:2005fw} vanish (except for the scaling dimension).},
but in the following discussion
we merely need it to be macroscopic.
By taking logarithm, differentiating with
respect to the spectral parameter $u$
and performing Fourier transform successively,
(\ref{BAE4bis}), (\ref{BAE7center}) give rise to
the following set of integral equations
\begin{eqnarray}\label{inteq4}
J_0(2gt)\Eqn{=}
e^{|t|}\hrho_4(t)+\hrho_4(t)
-4g^2 t \int_0^\infty dt'\hK_1(2gt,2gt')
\left[\hrho_2(t')+\hrho_6(t')\right],\\
\label{inteq6}
0\Eqn{=}
-e^{|t|}\hrho_6(t)+\hrho_6(t)
-4g^2 t \int_0^\infty dt'\hK_0(2gt,2gt')\hrho_4(t')\,.
\end{eqnarray}
The computation is almost parallel with that in \cite{Rej:2007vm},
where
Fourier transform of the density function is defined by
\begin{equation}
\hrho(t)=e^{-|t|/2}\int_{-\infty}^\infty e^{itu}\rho(u) du\,,
\end{equation}
and the integration kernels are given by
\begin{equation}
\hK_0(t,t')
=\frac{tJ_1(t)J_0(t')-t'J_0(t)J_1(t')}{t^2-{t'}^2},\quad
\hK_1(t,t')
=\frac{t'J_1(t)J_0(t')-tJ_0(t)J_1(t')}{t^2-{t'}^2}\,.
\end{equation}
$J_k(u)$ is the Bessel function of the first kind. 

Note that the first term of the r.h.s. of equations
(\ref{inteq4}), (\ref{inteq6})
comes from the growth of the mode number along the real axis,
while the second term comes from the scattering of the Bethe roots
of the same flavor. Observe that the relative signs of these two terms
are different in (\ref{inteq4}) and (\ref{inteq6}).
This is due to the fact that $u_{4,k}$ correspond to excitations
in the compact $\alg{so}(6)$ sector, while $u_{6,k}$ correspond
to the non-compact $\alg{so}(4,2)$ sector.

By eliminating $\hrho_2,\hrho_6$, one obtains a single integral
equation for $\hrho_4(t)$
\begin{equation}\label{inteq4bis}
J_0(2gt)=
(e^{|t|}+1)\hrho_4(t)
+4g^2 t \int_0^\infty dt'\hK_d(2gt,2gt')\hrho_4(t')\,,
\end{equation}
where the integration kernel reads
\begin{equation}\label{K_d}
\hK_d(t,t')=8g^2\int_0^\infty
dt'' \hK_1(t,2gt'')\frac{t''}{e^{t''}-1}\hK_0(2gt'',t')\,.
\end{equation}
We obtain the very kernel describing
the dressing phase proposed in \cite{Beisert:2006ez}.

\par\bigskip
The meaning of the result is as follows.
Let us first see what state we have at one loop
by taking the limit $g\to 0$ in
our starting equations (\ref{BAE4})--(\ref{BAE7}).
One immediately sees that
equations (\ref{BAE6}) and (\ref{BAE7})
decouple from (\ref{BAE4}).
In fact, they reduce to trivial equations via a duality transformation
and we are left only with
the simple Bethe ansatz equations for $u_{4,k}$
describing the $\alg{su}(2)$ Heisenberg spin-chain.
In particular, our solution with maximally filling
real $u_{4,k}$ corresponds to the anti-ferromagnetic state
of the $\alg{su}(2)$ chain.
In fact, (\ref{inteq4bis}) has the form of the 
continuous all-loop Bethe ansatz equation
for the $\alg{su}(2)$ anti-ferromagnetic state
\cite{Zarembo:2005ur,Rej:2005qt},
plus the integral term expressing
the contribution from the background stacks.

Next, let us consider the 
$\alg{su}(2)$ anti-ferromagnetic state
using the all-loop Bethe ansatz equations with the dressing factor
of Beisert, Eden and Staudacher \cite{Beisert:2006ez}
instead of introducing our background stacks.
In the thermodynamic limit, the same equation (\ref{inteq4bis})
appears \cite{Rej:2007vm}, but now the integral term comes
from the dressing factor. 
This shows that introduction of the background stacks
is equivalent to that of the dressing factor. 
In other words, the dressing phase is nothing but  the effective phase 
due to the existence of our background stacks, 
which provides an interesting clue to the physical
origin of the dressing factor. 

In this letter, we have focused on the $\alg{su}(2)$
anti-ferromagnetic state. Other states are also described similarly to
the hole excitations above the anti-ferromagnetic vacuum of the
$\alg{su}(2)$ Heisenberg spin-chain \cite{Sakai:2007ie}.

Now, one can argue that  there are two equivalent formulations also for 
planar ${\cal N}\!=\nobreak4$ Super Yang--Mills,
as discussed in the introduction:
one could start either from physical Bethe ansatz equations
with a trivial reference state, or from bare Bethe ansatz equations
with a non-trivial reference state. The former is derived from
the physical S-matrix involving the dressing factor, 
which is analogous to the Bethe ansatz formulation of quantum 
sigma-models \cite{Zamolodchikov:1978xm}.
The latter is derived from a bare R-matrix without the dressing factor, 
which is analogous to the lattice (spin-chain) realization of particle 
models.

Our result indicates  the possibility of the latter.
In this formulation, the fundamental R-matrix can be determined
by purely algebraic consideration \cite{Beisert:2005tm,Beisert:2006qh},
while the physical S-matrix is dynamically generated as the
scattering matrix of fundamental excitations over the Fermi surface.
The dressing phase is then regarded as the effective phase 
over the Fermi surface, or the ``physical vacuum'' with the stacks. 
To pursue this program,
there are still many questions open for further investigations.
One has to examine, for example, what are the allowed excitations, 
how a small number of excitations is described, and
how each Yang--Mills field is realized in terms of Bethe roots.
We leave detailed analysis for future publication \cite{Sakai:2007ie}.

\vspace{5ex}
\ni {\bf Note added:} \  
After the submission of this article we were informed
by A.~Rej, \hbox{M.~Staudacher} and S.~Zieme
that they were aware of similar results, which were
later presented in the revised version of \cite{Rej:2007vm}.
Despite the formal resemblance, our approach is
different from theirs in several ways, in particular conceptually,
and the open questions in \cite{Rej:2007vm}  are resolved in ours
\cite{Sakai:2007ie}.

%
\vspace{5ex}
\begin{center}
  {\bf Acknowledgments}
\end{center}
We would like to thank
H-Y.~Chen, N.~Dorey, V.~Kazakov, I.~Klebanov, T.~Klose,
C.~Kristjansen, M.~Martins, A.~Rej, D.~Serban,
M.~Staudacher, A.~Volovich, K.~Yoshida and S.~Zieme
for useful comments and discussions.
We are especially grateful to \hbox{M.~Shiroishi}
for collaborative discussions at the early stage of this project.
K.S. is very grateful to the
Laboratoire de Physique Th\'eorique de l'Ecole Normale Sup\'erieure,
the Department of Applied Mathematics and Theoretical Physics
at University of Cambridge,
and the String Theory Group at National Taiwan University
for their warm hospitality.
Research of K.S. is supported by
the Keio Gijuku Academic Development Funds.

\newpage
%
%
\def\thebibliography#1{\list
 {[\arabic{enumi}]}{\settowidth\labelwidth{[#1]}\leftmargin\labelwidth
  \advance\leftmargin\labelsep
  \usecounter{enumi}}
  \def\newblock{\hskip .11em plus .33em minus .07em}
  \sloppy\clubpenalty4000\widowpenalty4000
  \sfcode`\.=1000\relax}
 \let\endthebibliography=\endlist
%
%
\vspace{3ex}
\begin{center}
 {\bf References}
\end{center}
\par \vspace*{-2ex}

%

%

\end{document}